\documentclass[preprint,aps,amsmath,amssymb,12pt]{revtex4}
\usepackage{epsfig}
\usepackage{slashed}
\usepackage{graphicx}
\usepackage{subfigure}
\usepackage{epstopdf}
\usepackage{soul}
\usepackage{color}
\usepackage{booktabs}
\usepackage{multirow}
\usepackage{multirow,color}
\graphicspath{{PDF/}}
\begin{document}
\title{ Searching for the light leptophilic gauge boson $Z_x$ via four-lepton final states at the CEPC}
\author{Chong-Xing Yue}\email{cxyue@lnnu.edu.cn}
\author{Yan-Yu Li}\email{lyy3390@163.com }
\author{Mei-Shu-Yu Wang}\email{1404592974@qq.com }
\author{Xin-Meng Zhang}\email{398443768@qq.com }

\affiliation{
Department of Physics, Liaoning Normal University, Dalian 116029, China\\
Center for Theoretical and Experimental High Energy Physics, \\
Liaoning Normal University, China
\vspace*{1.5cm}}

\begin{abstract}
We investigate the possibility of detecting the leptophilic gauge boson $Z_x$ predicted by the $U(1)_{L_e-L_\mu}$ model via the processes $e^+e^-\rightarrow\ell^+\ell^-Z_x(Z_x\rightarrow \ell^+\ell^-)$ and $e^+e^-\rightarrow \ell^+\ell^-Z_x(Z_x\rightarrow \nu_\ell\bar{\nu_\ell})$ at the Circular Electron Positron Collider (CEPC) with the center of mass energy $\sqrt s=240$ GeV and the luminosity $\mathcal{L}=5.6$ $\mathrm{ab^{-1}}$. We give the expected sensitivities of the CEPC to the parameter space at $1\sigma$, $2\sigma$, $3\sigma$ and $5\sigma$ levels.

\end{abstract}


\maketitle
\section{Introduction}

The Circular Electron Positron Collider (CEPC) \cite{CEPCStudyGroup:2018ghi} is a particle physics research program of great scientific significance
and great potential. The concept of the CEPC is developed in the context of many international large colliders, such as the Large Hadron Collider
(LHC) at the European Center for Nuclear Research (CERN). Compared with their high energy consumption and cost, as well as the pressure of data
processing and storage, the CEPC has unique features and advantages. Firstly, the CEPC is an electron collider in which positrons and electrons collide with each other to produce high-energy particle events. Unlike hadron collisions, electron collisions produce particle events that are much clearer and more controllable, facilitating precise measurements and particle identification. Secondly, the CEPC plans to build a highly detailed detector that will be able to capture and record all the important information in particle collisions, providing physicists with a large amount of data to study the behavior of elementary particles. In addition, the CEPC will invest a great deal of effort in improving data processing and storage techniques to cope with the high density of collision data. Finally, the CEPC has a much brighter and cleaner experimental environment. Not only can standard model (SM) observables be studied with unprecedented precision, but also the precision of many electroweak observables will be improved by an order of magnitude or more. So the CEPC offers an unmatched opportunity for precision measurements and searches for beyond the standard model (BSM) physics.

Among the many new physics (NP) scenarios, there is a class of models that predict the existence of leptophilic gauge boson $Z_x$, this kind of new
neutral gauge boson arises due to the extension of a group in the Standard Model with the $U(1)_{L_x-L_y}$ for $x, y \in \{e, \mu, \tau$\} \cite{Foot:1990mn,He:1991qd,Foot:1994vd}. The global symmetry $U(1)_{L_x-L_y}$ can be introduced to the SM, which is anomaly-free without any
additional particle \cite{Buchmuller:1991ce,Marshak:1979fm}. When the $U(1)_{L_x-L_y}$ gauge symmetry is spontaneously broken, the leptophilic gauge
boson $Z_x$ gains mass. This class of models can be a good solution to solve some problems in the SM, such as the neutrino mass and mixing problem \cite{Baek:2015mna,Heeck:2011wj,Biswas:2016yan}, the dark matter dark energy problem \cite{Biswas:2016yan,Patra:2016shz,Arcadi:2018tly,Altmannshofer:2016jzy}, and the muon anomalous magnetic moment problem
\cite{Gninenko:2001hx,Ma:2001md,Baek:2001kca}. In our work, we discuss the possibility of probing this class of the leptophilic gauge boson $Z_x$ at
the CEPC.

The study of the leptophilic gauge boson $Z_x$ is an important step in the exploration of NP. The $Z_x$ boson can be produced at current collider
experiments. For example, at the LHC, the leptophilic gauge boson $Z_x$ is mainly produced via the Drell-Yan process, where the $Z_x$ is radiated from the final-state leptons, the constraints on the $Z_x$ boson can be given via the processes $pp\rightarrow 4\ell, 3\ell+E_T^{miss}, 2\ell+E_T^{miss}$, or $1\ell+E_T^{miss}$ \cite{Medina:2021ram,CMS:2017dzg,CMS:2017moi,Drees:2018hhs}. At the KEKB collider, the leptophilic gauge boson $Z_x$ is produced via the process $e^+e^-\rightarrow \mu^+\mu^-Z_x(Z_x\rightarrow \mu^+\mu^-)$ in the framework of the $U(1)_{L_\mu-L_\tau}$ model in the small mass range $M_{Z_x}<10$ GeV \cite{Belle:2021feg}. The processes $e^+e^-\rightarrow Z_x\rightarrow\ell^+\ell^-$ or $q\bar{q}$ \cite{ALEPH:2013dgf} and $e^+e^-\rightarrow \gamma Z_x(\rightarrow\ell^+\ell^-, \mu^+\mu^-)$ \cite{BaBar:2014zli} can also be used to search for the $Z_x$ boson at the LEP and BABAR. Most of the LHC (and Tevatron) bounds coming from resonance searches do not directly apply to such a neutral leptophilic sector, the relevant collider constraints of the $U(1)_{L_e-L_\mu}$ model mainly come from LEP, and are generally much weaker than the direct LHC constraints applicable for hadrophilic resonances \cite{Dasgupta:2023zrh}, so the future $e^+e^-$ colliders are uniquely capable of probing the leptophilic gauge boson $Z_x$ to unprecedented mass and coupling values. Refs. \cite{He:2017zzr, Liu:2019ogn} have studied the sensitivity of the process $e^+e^-\rightarrow Z_x \gamma$ to explore the leptophilic gauge boson $Z_x$ in the future $e^+e^-$ colliders. In general, properties of any new particle can be studied via different processes even if at the same collider experiments. Furthermore, we find that there are few studies to search for the gauge boson $Z_x$ predicted by the $U(1)_{L_e-L_\mu}$ model via four-lepton final state processes at the future $e^+e^-$ colliders, so we propose to search for this kind of leptophilic gauge boson $Z_x$  via the processes $e^+e^-\rightarrow \ell^+\ell^-Z_x(Z_x\rightarrow \ell^+\ell^-)$ and $e^+e^-\rightarrow \ell^+\ell^-Z_x(Z_x\rightarrow \nu_\ell\bar{\nu_\ell})$ at the 240 GeV CEPC. We expect these processes to give better sensitivities in the certain mass range.

The paper is organized as follows. In Section \uppercase\expandafter{\romannumeral2} we will briefly introduce the $U(1)_{L_e-L_\mu}$ model and summarize the constraints of existing experiments on the model. Based on the details of the analysis of the $Z_x$ signal processes $e^+e^-\rightarrow \ell^+\ell^-Z_x(Z_x\rightarrow\ell^+\ell^-)$ and $e^+e^-\rightarrow \ell^+\ell^-Z_x(Z_x\rightarrow \nu_\ell\bar{\nu_\ell})$ and the relevant SM backgrounds, sensitivity projections of the CEPC to the $U(1)_{L_e-L_\mu}$ model parameter space are presented, and compared with other experimental results in Section \uppercase\expandafter{\romannumeral3}. Finally the conclusion and discussion are given in Section \uppercase\expandafter{\romannumeral4}.

\section{The $U(1)_{L_e-L_\mu}$ model}

\begin{table}[!htb]
\begin{center}
\caption{Lepton charges corresponding to the $U(1)_{L_x-L_y}$ models.}
\label{tab1}
\begin{tabular}{|c|c|c|c|}\hline
\multirow{2}*{~~~Model~~~}             &\multicolumn{3}{|c|}{Charge} \\ \cline{2-4}
~                          &~~~$e,\nu_e$~~~    &$\mu, \nu_\mu$    &$\tau, \nu_\tau$        \\  \hline
~~~$L_e-L_\mu$~~~             &~~~$1$~~~       &$-1$   &$0$          \\ \hline
~~~$L_e-L_\tau$~~~             &~~~$1$~~~       &$0$   &$-1$ \\ \hline
~~~$L_\mu-L_\tau$~~~           &~~~$0$~~~       &$1$   &$-1$    \\ \hline
~~~$L_e-\frac{1}{2}(L_\mu+L_\tau)$~~~             &~~~$1$~~~       &$-\frac{1}{2}$   &$\frac{1}{2}$         \\ \hline
~~~$L_e+2(L_\mu+L_\tau)$~~~           &~~~$1$~~~       &$2$   &$2$          \\ \hline
\end{tabular}
\end{center}
\end{table}
The $U(1)_{L_x-L_y}$ model \cite{Foot:1990mn,He:1991qd,Foot:1994vd} is composed of the SM gauge group $SU(3)_C\otimes SU(2)_L\otimes U(1)_Y$ expanding a $U(1)_{L_x-L_y}$ group without introducing an anomaly, this surprising feature is the main motivation considered here. For convenience, in Table \ref{tab1}, we list the lepton charges for the $U(1)_{L_x-L_y}$ models, $e$, $\mu$, and $\tau$ are three generations of charged leptons, $\nu_e$, $\nu_\mu$, and $\nu_\tau$ represent the corresponding left-handed neutrinos, respectively.

The part of Lagrangian of the $U(1)_{L_e-L_\mu}$ model can be written as
\begin{eqnarray}
\mathcal{L}(Z_x)&=&-g'Z_x{^{\alpha}}[Q_e(\bar{e}\gamma_\alpha e+\bar{\nu_e}\gamma_\alpha P_L \nu_e)+Q_\mu(\bar{\mu}\gamma_\alpha
\mu+\bar{\nu_\mu}\gamma_\alpha P_L \nu_\mu)\nonumber\\
&&+Q_\tau(\bar{\tau}\gamma_\alpha \tau+\bar{\nu_\tau}\gamma_\alpha P_L
\nu_\tau)]-\frac{1}{4}Z_x{_{\mu\nu}}Z_x{^{\mu\nu}}+\frac{1}{2}m_{Z_x}^2Z_x{^\mu} Z_x{_\mu},
\end{eqnarray}
where the gauge coupling constant is denotes as $g'$, $P_L=\frac{1}{2}(1-\gamma_5)$ is the left chirality projector, $Q_e$,
$Q_\mu$, and $Q_\tau$ respectively correspond to the charges of lepton of three generations in the $U(1)_{L_e-L_\mu}$, and the $Z_x-$ field strength tensor can be written as
\begin{eqnarray}
Z_x{_{\mu\nu}}=\partial_{\mu} Z_x{_\nu}-\partial_{\nu} Z_x{_\mu}.
\end{eqnarray}

Before we discuss the experimental constraints on the gauge boson $Z_x$, let us present the decays of the $Z_x$ boson. The partial decay width of $Z_x\rightarrow \ell^+\ell^-(\nu_\ell\bar{\nu_\ell})$ for a single flavor lepton is given by
\begin{eqnarray}
\Gamma(Z_x\rightarrow \ell^+ \ell^-)=\frac{(g'Q_\ell)^2M_{Z_x}}{12\pi}(1+\frac{2m_\ell^2}{M^2_{Z_x}})\sqrt{1-\frac{4m_\ell^2}{M_{Z_x}^2}},
\end{eqnarray}
\begin{eqnarray}
 \Gamma(Z_x\rightarrow \nu_\ell\bar{\nu_\ell})= \frac{(g'Q_\ell)^2M_{Z_x}}{24\pi}.
\end{eqnarray}
In the $U(1)_{L_e-L_\mu}$ model, the gauge boson $Z_x$ can only couple to two flavor leptons, so the decay channels of the $Z_x$ boson are as follows
\begin{eqnarray}
Z_x\rightarrow e^+ e^-, Z_x\rightarrow \mu^+ \mu^-, Z_x\rightarrow \nu_e\bar{\nu_e}, Z_x\rightarrow \nu_\mu\bar{\nu_\mu}.
\end{eqnarray}
Since $M_{Z_x} \gg M_\ell$, we can neglect the mass of the lepton in Eq. (3), which gives the total width of the gauge boson $Z_x$ as
\begin{eqnarray}
\Gamma_{Z_x}\simeq\frac{g'^2}{4\pi}M_{Z_x}.
\end{eqnarray}

There are two possible ways to discover the $Z_x$ boson. On the one hand, the $Z_x$ boson is heavy at the current energy, and we would need a higher
energy to find it. On the other hand, it may be that the $Z_x$ mass is very light and the coupling to the particles in the SM is weak (similar to the
search for the Higgs boson), so the people search for it by directly or indirectly production at the future colliders. In our work, we prefer the latter. When the boson $Z_x$ has a light mass, the $\tau$ mass is heavy and unstable, we mainly consider that the $Z_x$ boson couples only to the $e$ and $\mu$ subsets and their corresponding neutrinos in the $U(1)_{L_e-L_\mu}$ model. Some existing constraints on the leptophilic gauge boson mass $M_{Z_x}$ and coupling $g'$ in the $U(1)_{L_e-L_\mu}$ model are summarized by Ref. \cite{Chun:2018ibr}. The LEP bounds give the most stringent bounds in the larger mass range $M_{Z_x} \leq 10^3$ GeV at $1\sigma(2\sigma)$ via $e^+e^-\rightarrow \ell^+\ell^-$ processes, CMS investigated the final state $4\mu$ for the case that all muons originate from the decay of an(almost) on-shell $Z$ boson, offering good sensitivity for $10$ GeV $ <M_{Z_x} < 60$ GeV. The strongest constraints on the coupling $g'$ with the $10-60$ GeV mass range comes from the LHC at $95\%$ confidence level(CL), the $g'$ can be as low as $2\times10^{-2}$ \cite{Dasgupta:2023zrh,Drees:2018hhs}. The production of a muon-antimuon pair in the scattering of muon neutrinos in the Coulomb field of a target nucleus gives the strong bound, e.g., neutrino trident production \cite{Altmannshofer:2014pba,Belusevic:1987cw}. And a combination of measurements of the trident cross section from CHARM-\uppercase\expandafter{\romannumeral2} \cite{CHARM-II:1990dvf}, CCFR \cite{CCFR:1991lpl} and NuTeV \cite{NuTeV:1998khj} imposes a bound of $g'\lesssim1.9\times10^{-3}\mathrm{M_{Z_x}/GeV}$ on the $U(1)_{L_e-L_\mu}$ \cite{Dasgupta:2023zrh}. The sensitivities from $(g-2)_e$ and $(g-2)_\mu$ on $g'$ in the $U(1)_{L_e-L_\mu}$ model are also considered, their results are respectively in the range $0.2-1$ and $4\times10^{-2}-1$ \cite{Dasgupta:2023zrh,Pospelov:2008zw,Laine:2022ytc,Aoyama:2019ryr,Morel:2020dww,Parker:2018vye,Muong-2:2021ojo,Muong-2:2023cdq,Muong-2:2006rrc,Aoyama:2020ynm}. So we propose the process $e^+e^-\rightarrow \ell^+\ell^-Z_x(Z_x\rightarrow \ell^+\ell^-$ or $\nu_\ell\bar{\nu_\ell})$ in the $U(1)_{L_e-L_\mu}$ model with $10$ GeV $\leq M_{Z_x} \leq 60$ GeV to go further in the search for expected sensitivities of the $Z_x$ boson at the$\sqrt{s}=240$ GeV CEPC.

\section{searching for $Z_x$ at the CEPC}

The main Feynman diagrams of the signal process $e^+e^-\rightarrow \ell^+\ell^-Z_x(Z_x\rightarrow \ell^+\ell^-$ or $\nu_\ell\bar{\nu_\ell})$ are shown in Figure 1,
\begin{figure}
  \centering{
  \subfigure[]{\includegraphics[width=0.31\textwidth]{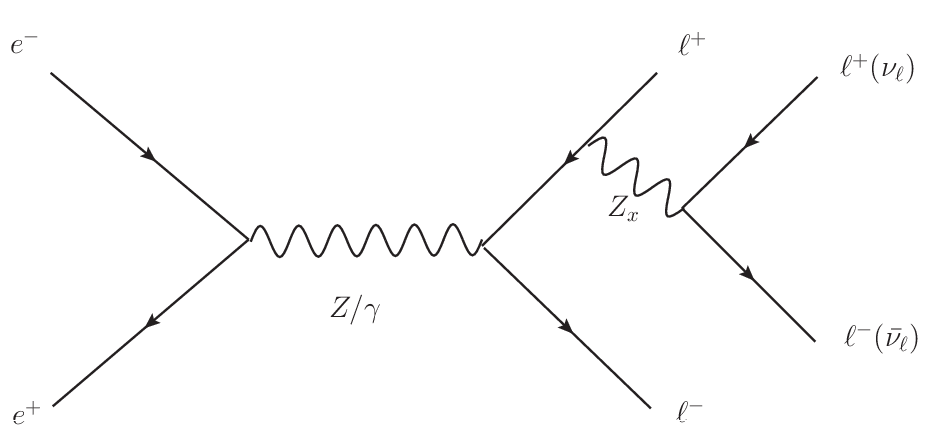}}
  \subfigure[]{\includegraphics[width=0.31\textwidth]{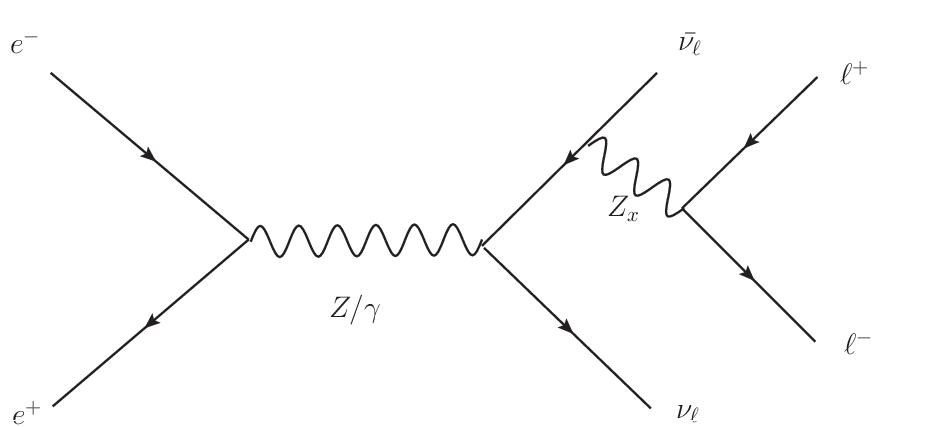}}
  \subfigure[]{\includegraphics[width=0.31\textwidth]{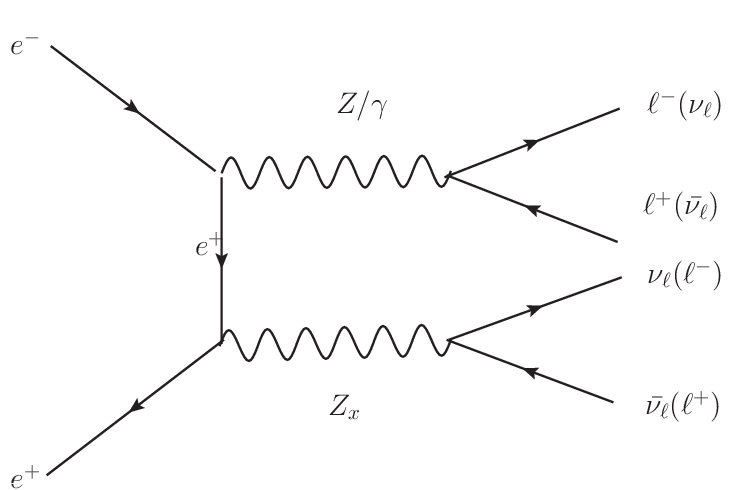}}
  \caption{The main Feynamn diagrams for the process $e^+e^-\rightarrow \ell^+\ell^-Z_x(Z_x\rightarrow \ell^+\ell^-$ or $\nu_\ell\bar{\nu_\ell})$ within $\ell \in \{e, \mu\}$.}
  \label{FIG1}}
\end{figure}
which can be expanded into the following four processes, $e^+e^-\rightarrow e^+e^-Z_x(Z_x\rightarrow e^+e^-)\rightarrow e^+e^-e^+e^-$,
$e^+e^-\rightarrow \mu^+\mu^-Z_x(Z_x\rightarrow \mu^+\mu^-)\rightarrow \mu^+\mu^-\mu^+\mu^-$, $e^+e^-\rightarrow e^+e^-Z_x(Z_x\rightarrow
\nu_e\bar{\nu_e})\rightarrow e^+e^-\nu_e\bar{\nu_e}$, and $e^+e^-\rightarrow \mu^+\mu^-Z_x(Z_x\rightarrow \nu_\mu\bar{\nu_\mu})\rightarrow
\mu^+\mu^-\nu_\mu\bar{\nu_\mu}$. In Figure 2, we give the cross sections of four signaling processes and the corresponding backgrounds, the numerical
results for the cross sections are imposed on the basic cuts. We make the transverse momenta of the leptons $P_T(\ell)$ greater than 10 GeV and the
absolute value of the lepton pseudorapidity $\eta_\ell$ needs to be less than 2.5. These basic cuts are then summed up as
\begin{eqnarray}
P_T(\ell)>10~\mathrm{GeV},&&\mid\eta_\ell\mid<2.5.
\end{eqnarray}

When the leptophilic gauge boson $Z_x$ decays to a pair of neutrinos, the beam polarizations can help further suppressing the SM backgrounds to
enhance the signals \cite{Ge:2023wye}. So in the right panel of Figure 2, we show the polarized cross sections of the $e^+e^-\rightarrow
\ell^+\ell^-\nu_\ell\bar{\nu_\ell}$ processes with the beam polarization configurations $(P_{e^+}, P_{e^-})=(-30\%, +80\%)$. The solid lines
represent the signal cross sections and the dashed lines represent the SM background cross sections. The cross sections range of
signal processes $e^+e^-\rightarrow e^+e^-\nu_e\bar{\nu_e}$ and $e^+e^-\rightarrow \mu^+\mu^-\nu_\mu\bar{\nu_\mu}$ for $10$ GeV $\leq M_{Z_x}\leq 60$
GeV are $3.24\times10^{-4}$ $-$ $1.6\times10^{-3}$ pb and $1.7\times10^{-4}$ $-$ $2.24\times10^{-4}$ pb with $g'=0.01$ $\mathrm{GeV^{-1}}$. For each of the above two processes, the background cross sections are 0.04823 pb and 0.03663 pb, respectively.
\begin{figure}
  \centering{
  \subfigure[]{\includegraphics[width=0.49\textwidth]{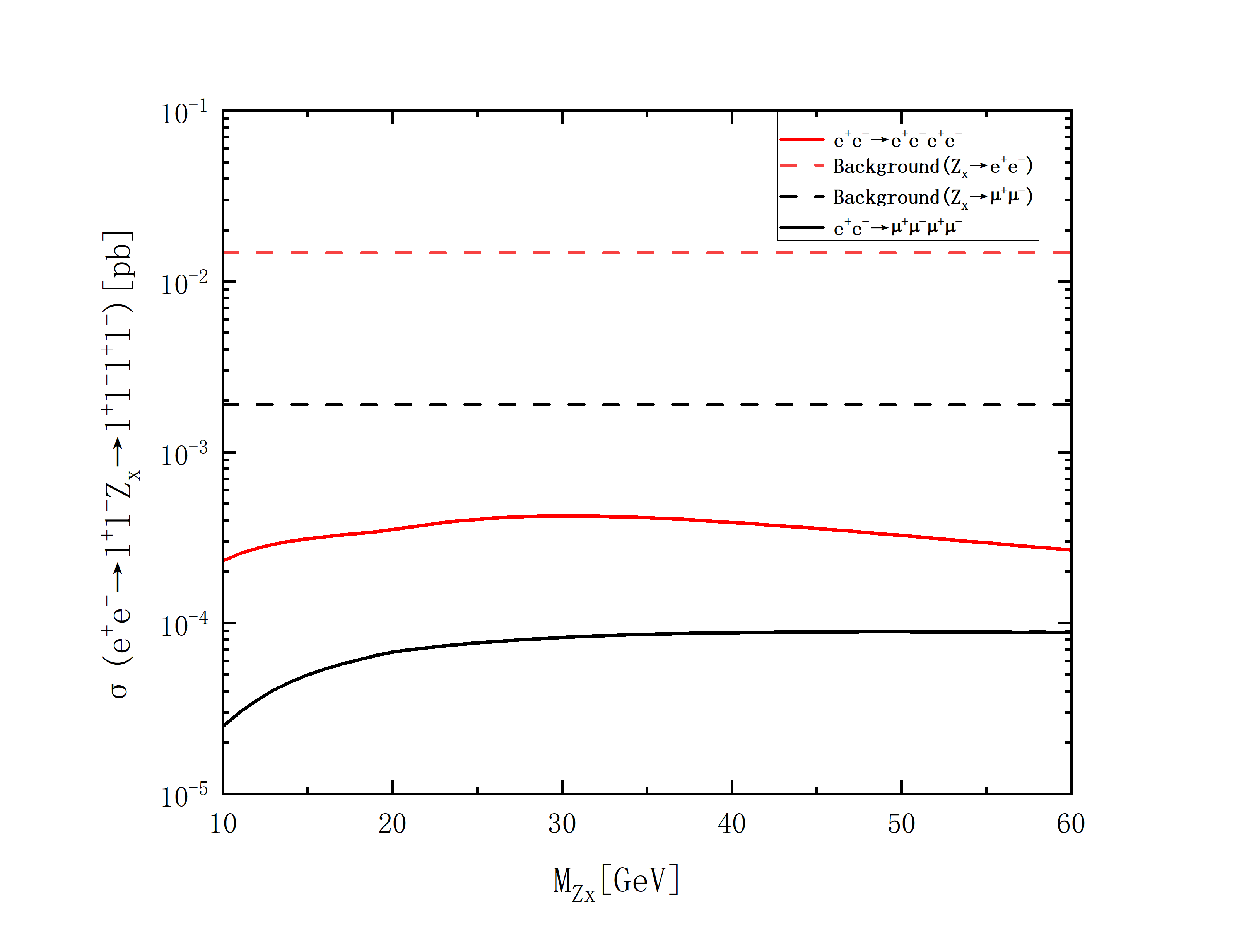}}
  \subfigure[]{\includegraphics[width=0.49\textwidth]{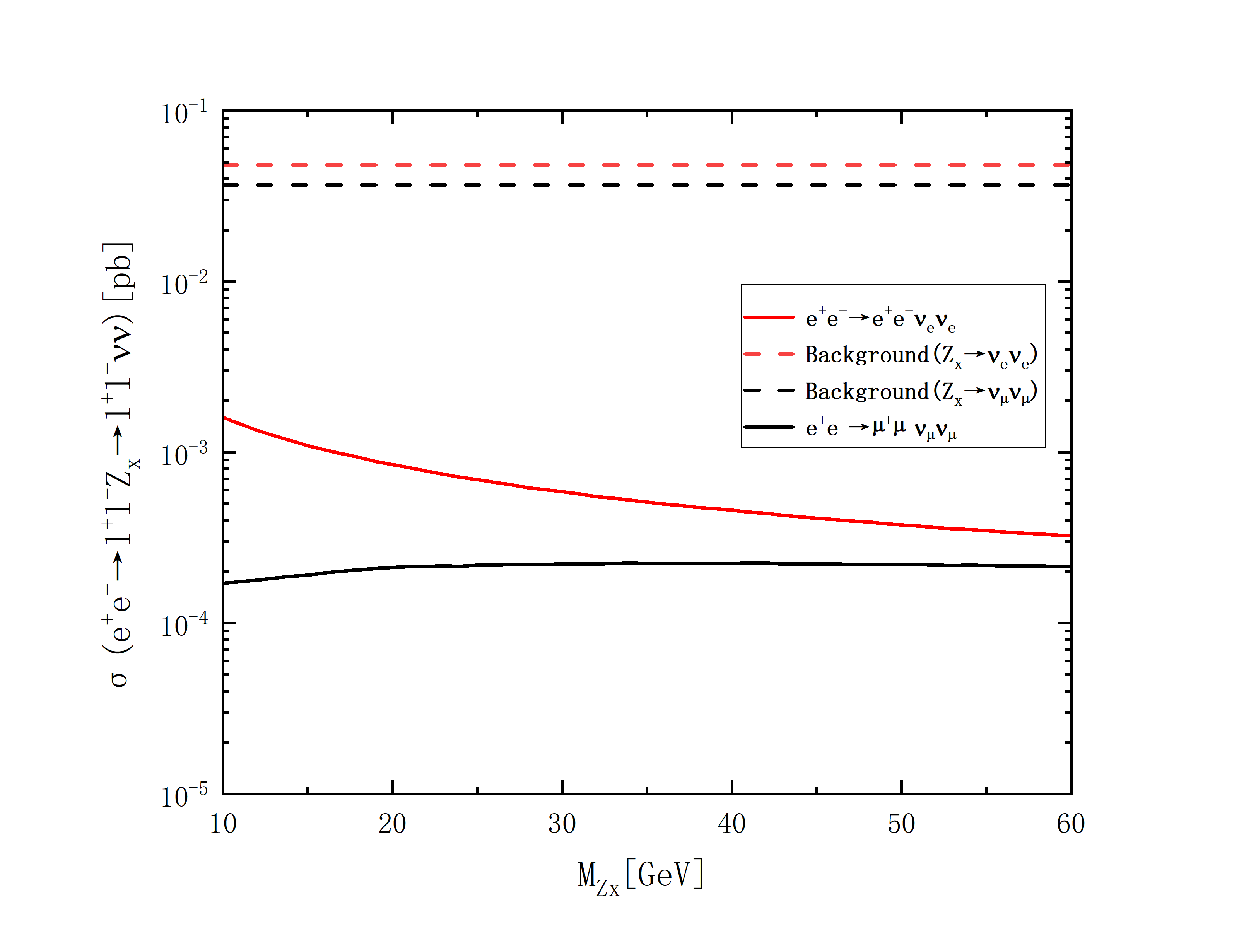}}
  \caption{The cross sections of the signal and background processes as functions of the mass $Z_x$ when the coupling limits $g'=0.01$ $\mathrm{GeV^{-1}}$.}
  \label{FIG2}}
\end{figure}

The left panel shows the $Z_x$ boson decaying to a pair of leptons, we consider the effect of polarization on the processes $Z_x\rightarrow \ell^+\ell^-$, but the variations in the cross sections are not significant, so we do not impose the beam polarizations on the cross sections. The solid-red and solid-black lines represent the cross sections of the signal processes $e^+e^-\rightarrow e^+e^-e^+e^-$ and $e^+e^-\rightarrow \mu^+\mu^-\mu^+\mu^-$, respectively, the numerical results are $2.31\times10^{-4}$ $-$ $4.24\times10^{-4}$ pb and $2.49\times10^{-5}$ $-$ $8.91\times10^{-5}$ pb in the mass range $10$ GeV $\leq M_{Z_x}\leq 60$ GeV when the coupling constant $g'=0.01$ $\mathrm{GeV^{-1}}$. The dashed-red and dashed-black lines represent the cross sections of the background processes $e^+e^-\rightarrow e^+e^-e^+e^-$(0.01477 pb) and $e^+e^-\rightarrow \mu^+\mu^-\mu^+\mu^-$(0.001899 pb), respectively. The cross sections of signals in the parameter region are smaller than the cross sections of corresponding SM backgrounds.

Next, in order to simulate the signals, we use FeynRules \cite{Alloul:2013bka} to produce a model file output in UFO format. Then all signal and
background events were simulated using MadGraph5 \cite{Alwall:2014hca}, the parton shower and hadronization are carried out with Pythia8
\cite{Sjostrand:2007gs}, while the detector simulation is performed inside MadAnalysis5 \cite{Conte:2012fm} and Delphes3 \cite{deFavereau:2013fsa}.
In our analysis, we generate, in each case, 10k signal events in an interval where the mass of $Z_x$ increases in order from 10 GeV to 60 GeV and
500k events for backgrounds.

\subsection{The visible decay channel $Z_x\rightarrow \ell^+\ell^-$}

In order to further improve event selection, the signal and background distributions of the angular separation $\triangle R$ between two muons which
is defined as $\triangle R=\sqrt{(\triangle\phi)^2+(\triangle\eta)^2}$ and invariant masses $M(\mu^+, \mu^-)$ are shown in Figure 3.\begin{figure}
  \centering{
  \subfigure[]{\includegraphics[width=0.49\textwidth]{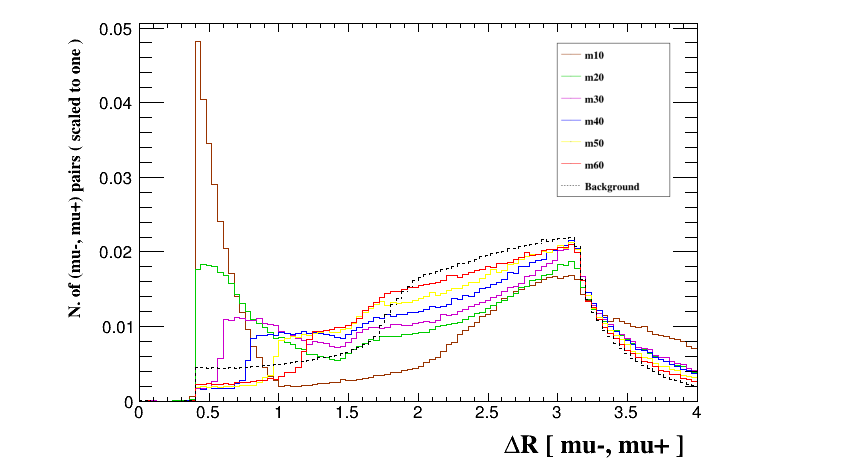}}
  \subfigure[]{\includegraphics[width=0.49\textwidth]{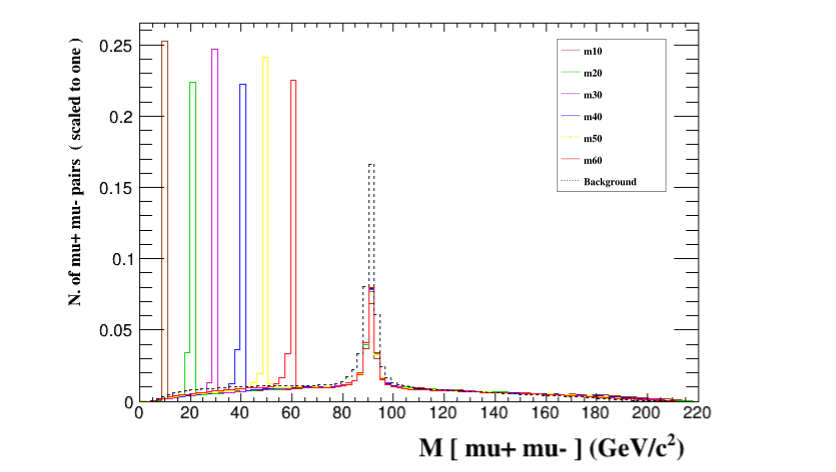}}
  \caption{Normalized distributions of $\triangle R$ (a) and $M(\mu^+, \mu^-)$ (b) from the signal and background events for different $M_{Z_x}$
  benchmark points for the process $e^+e^-\rightarrow \mu^+\mu^-\mu^+\mu^-$ at the CEPC with $\sqrt s=240$ GeV and $\mathcal{L}=5.6$ $\mathrm{ab^{-1}}$.}
  \label{FIG3}}
\end{figure}
\begin{table}[!htb]
\begin{center}
\caption{The improved cuts for the process $e^+e^-\rightarrow \mu^+\mu^-\mu^+\mu^-$.}
\label{tab2}
\vspace*{0.5cm}
\begin{tabular}{ c c c }\hline
\multirow{2}*{~~~Cut~~~}             &\multicolumn{2}{c}{Mass} \\ \cline{2-3}
~                          &~~~$10\mathrm{GeV}\leq M_{Z_x}\leq30\mathrm{GeV}$~~~    &$30\mathrm{GeV}< M_{Z_x}\leq60\mathrm{GeV}$         \\  \hline
~~~Cut1~~~             &~~~$\triangle R>0.5$~~~       &$\triangle R>0.7$          \\
~~~Cut2~~~             &~~~$M(\mu^+, \mu^-)-M_{Z_x}\leq5$~~~       &$M(\mu^+, \mu^-)-M_{Z_x}\leq5$          \\ \hline
\end{tabular}
\end{center}
\end{table}We can see that the background and signal have very distinctive characteristics. In particular, for the distribution of invariant masses
$M(\mu^+, \mu^-)$, the peaks in $M(\mu^+, \mu^-)$ still denounce the presence of signals making the distinction against the smooth background an easy
task. We select $M(\mu^+, \mu^-)-M_{Z_x}\leq5$. $\triangle R$ is greater than $0.5$ for $Z_x$ mass from 10 GeV to 30 GeV, and greater than 0.7 when the $Z_x$ mass is in the mass range $30$ $-$ $60$ GeV for the process $e^+e^-\rightarrow \mu^+\mu^-\mu^+\mu^-$. Based on the characteristics of the kinematic distributions, the selected cuts are listed in Table \ref{tab2}. After these improved cuts are applied, the SM background is significantly depressed. We take a signal benchmark point every 10 GeV in the $10$ $-$ $60$ GeV mass interval, and display the cross sections of the signal and background after applying the above selection cuts for these benchmark points for $g'=0.01$ $\mathrm{GeV^{-1}}$ at the 240 GeV CEPC with $\mathcal{L}=5.6$ $\mathrm{ab^{-1}}$ in Table \ref{tab3}.\begin{figure}
  \centering
  \includegraphics[width=0.6\textwidth]{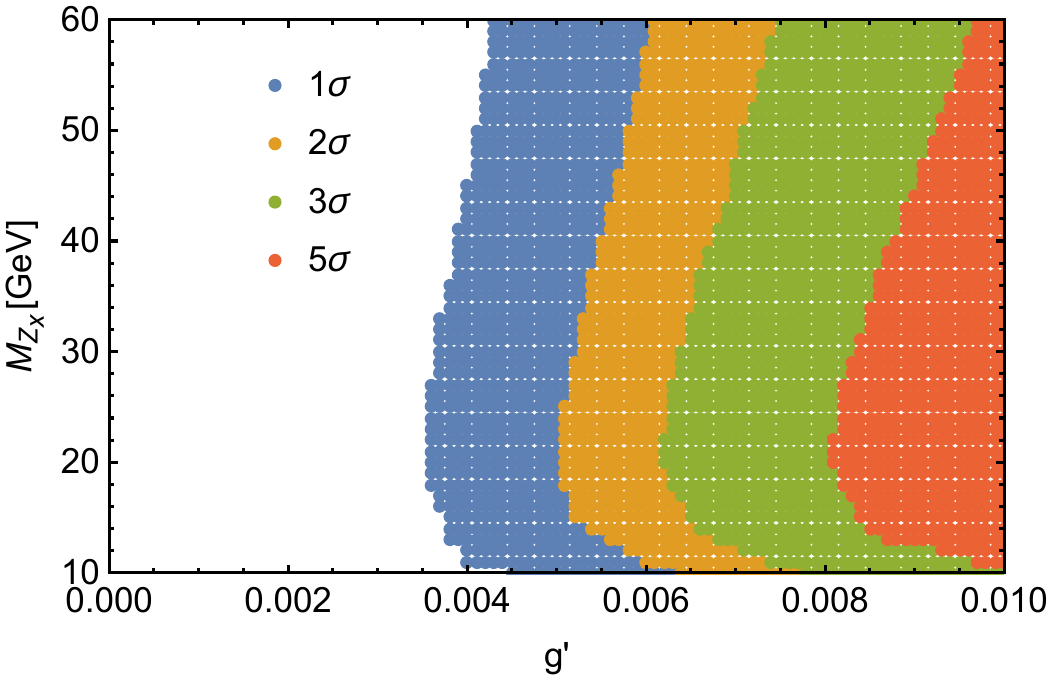}\\
  \caption{The $1\sigma$, $2\sigma$, $3\sigma$ and $5\sigma$ regions for the process $e^+e^-\rightarrow \mu^+\mu^-\mu^+\mu^-$ at the CEPC with $\sqrt s = 240$ GeV and $\mathcal{L}=5.6$ $\mathrm{ab^{-1}}$ in the $g'$-$M_{Z_x}$ plane.}\label{FIG 4}
\end{figure}
\begin{table}[!htb]
\begin{center}
\caption{The cross sections of the signal and background after imposing the improved cuts for $g'=0.01$ $\mathrm{GeV^{-1}}$ at the CEPC with $\sqrt s=240$ GeV and $\mathcal{L}=5.6$ $\mathrm{ab^{-1}}$ for the process $e^+e^-\rightarrow \mu^+\mu^-\mu^+\mu^-$.}
\label{tab3}
\begin{tabular}{c c c c c}\hline
  \multicolumn{5}{c}{Cross sections for signal(background) (fb)} \\ \hline
    $M_{Z_x}$ (GeV)  &~~Basic cuts ~~ & ~~Cut1~~&~~Cut2~~&~~SS~~\\ \hline
     10    & $2.4852\times10^{-2}$ ~~&~~ $2.4804\times10^{-2}$ ~~&~~ $2.3443\times10^{-2}$ ~~&~~4.7640\\
     ~     &(1.899)                     &(1.894)                     &(0.112)                     &~     \\
     20    & $6.7516\times10^{-2}$~~ & ~~$6.7369\times10^{-2}$~~ &~~ $6.3763\times10^{-2}$ ~~&~~7.4520\\
     ~     &(1.899)                     &(1.894)                     &(0.346)                     &~     \\
     30    & $8.2532\times10^{-2}$ ~~& ~~$8.2393\times10^{-2}$~~ &~~ $7.8353\times10^{-2}$ ~~&~~7.1090\\
     ~     &(1.899)                     &(1.894)                     &(0.601)                     &~     \\
     40    & $8.7872\times10^{-2}$ ~~& ~~$8.7711\times10^{-2}$ ~~& ~~$8.3870\times10^{-2}$ ~~&~~6.4464\\
     ~     &(1.899)                     &(1.894)                     &(0.864)                     &~     \\
     50    & $8.8983\times10^{-2}$~~ &~~ $8.8865\times10^{-2}$ ~~&~~ $8.5551\times10^{-2}$~~ &~~5.8548\\
     ~     &(1.899)                     &(1.894)                     &(1.109)                     &~     \\
     60    & $8.8160\times10^{-2}$~~ &~~ $8.8096\times10^{-2}$ ~~&~~ $8.5297\times10^{-2}$~~ &~~5.3669\\
     ~     &(1.899)                     &(1.894)                     &(1.325)                     &~     \\
     \hline
\end{tabular}
\end{center}
\end{table}We also show the statistical significance (SS) in the last column of Table \ref{tab3}, which defined as $\mathrm{SS} = \mathrm{S}/\sqrt{\mathrm{S}+\mathrm{B}}$, where $\mathrm{S}$ represents the number of signal events and $\mathrm{B}$ represents the number of background events. The $1\sigma$, $2\sigma$, $3\sigma$ and $5\sigma$ regions in the $g'$-$M_{Z_x}$ plane are plotted in Figure 4. The expected bounds on $g'$ can reach $6.2\times10^{-3}$ ($8.1\times10^{-3}$) $\mathrm{GeV^{-1}}$ at $3\sigma$ $(5\sigma)$ levels. Compared to the same signal process for the mass $M_{Z_x}<10$ GeV, the Ref. \cite{Belle:2021feg} gives the upper limit on $g'$ at SS $=1\sigma$ level, but we can give the SS at $3\sigma$ $(5\sigma)$ levels for the mass range $10-60$ GeV. Thus, the CEPC has potential to discover the $Z_x$ boson in the consider mass range.

\begin{figure}
  \centering{
  \subfigure[]{\includegraphics[width=0.49\textwidth]{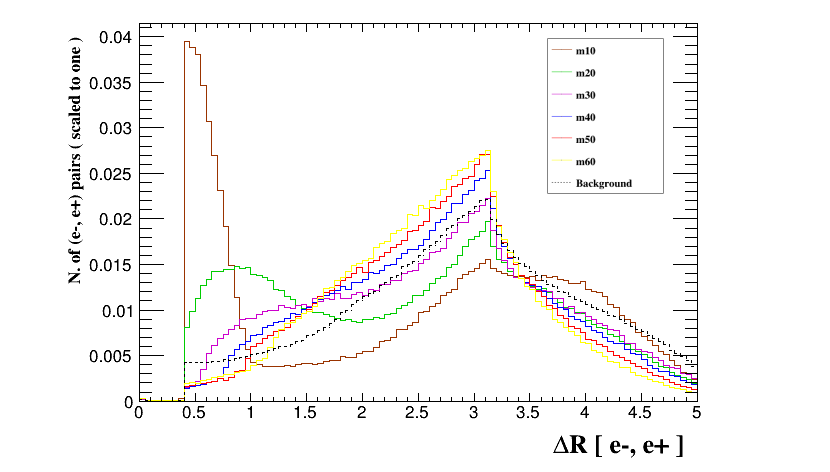}}
  \subfigure[]{\includegraphics[width=0.49\textwidth]{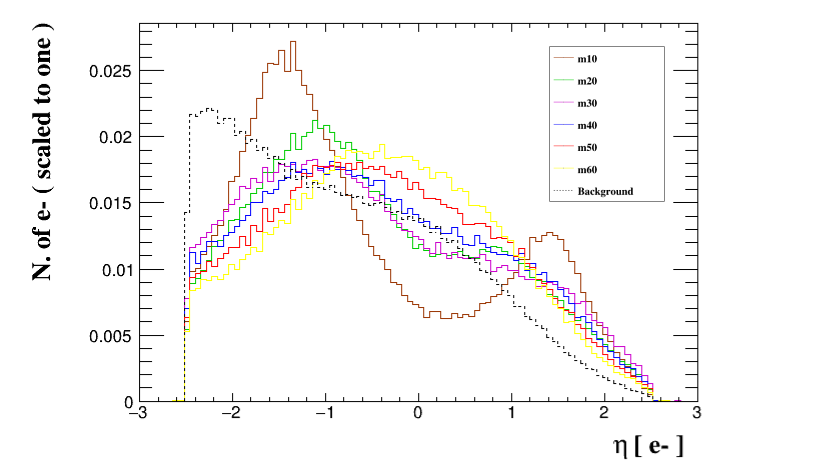}}\\
  \subfigure[]{\includegraphics[width=0.49\textwidth]{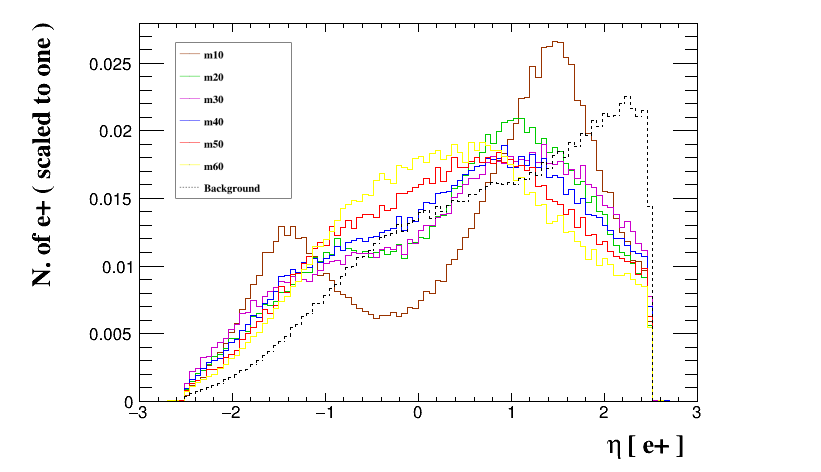}}
  \subfigure[]{\includegraphics[width=0.49\textwidth]{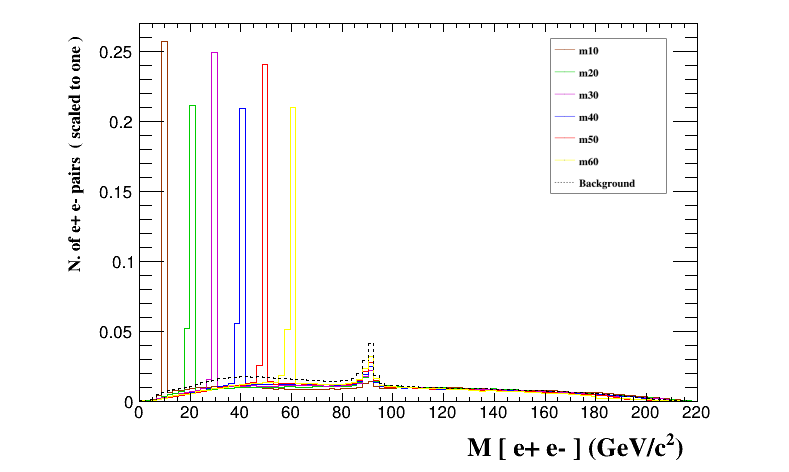}}
  \caption{Normalized distributions of $\triangle R$ (a), $\eta_{e^-}$ (b), $\eta_{e^+}$ (c) and  $M(e^+, e^-)$ (d) from the signal and background
  events for different $M_{Z_x}$ benchmark points for the process $e^+e^-\rightarrow e^+e^-e^+e^-$ at the CEPC with $\sqrt s=240$ GeV and
  $\mathcal{L}=5.6$ $\mathrm{ab^{-1}}$.}
  \label{FIG 5}}
\end{figure}
When the gauge boson $Z_x$ decays into a pair of electrons, the kinematic distributions of the signal process $e^+e^-\rightarrow e^+e^-e^+e^-$, $\triangle R$, $\eta_{e^-}$, $\eta_{e^+}$, and $M(e^+, e^-)$ are shown in Figure 5. The mass of $Z_x$ is greater than 40 GeV, the distribution of the peak of $\eta_{e^-}$ and $\eta_{e^+}$ are shifted, so we will divide the mass range into two segments of $10-40$ GeV and $40$ $-$ $60$ GeV when we select the effective cuts, ultimately we summarized the specific cuts in Table \ref{tab4}. After applying improved cuts, the cross sections of the signal and the background are shown in Table \ref{tab5}. We also give the regions of SS at $1\sigma$, $2\sigma$, $3\sigma$ and $5\sigma$ levels in Figure 6, as can be seen from the figure, the sensitivity projections of $Z_x$ become weaker with increasing mass and there is a significant dip at $M_{Z_x}=30$ GeV with $g'=5\times10^{-3}$ $\mathrm{GeV^{-1}}$. By comparing the above two processes, the four-electron final state is more sensitive to discover the $Z_x$ boson.
\begin{table}[!htb]
\begin{center}
\caption{The improved cuts for the process $e^+e^-\rightarrow e^+e^-e^+e^-$.}
\label{tab4}
\vspace*{0.5cm}
\begin{tabular}{ c c c }\hline
\multirow{2}*{~~~Cut~~~}             &\multicolumn{2}{c}{Mass} \\ \cline{2-3}
~                          &~~~$10\mathrm{GeV}\leq M_{Z_x}\leq40\mathrm{GeV}$~~~    &$40\mathrm{GeV}< M_{Z_x}\leq60\mathrm{GeV}$         \\  \hline
~~~Cut1~~~             &~~~$\triangle R>0.7$~~~       &$\triangle R>1$          \\
~~~Cut2~~~             &~~~$\eta_{e^-} >-1.4$~~~       &$\eta_{e^-} >-1.1$          \\
~~~Cut3~~~             &~~~$\eta_{e^+} <1.4$~~~       &$\eta_{e^+} <1.1$          \\
~~~Cut4~~~             &~~~$M(e^+, e^-)-M_{Z_x}\leq5$~~~       &$M(e^+, e^-)-M_{Z_x}\leq5$          \\ \hline
\end{tabular}
\end{center}
\end{table}\begin{table}[!htb]
\begin{center}
\caption{The cross sections of the signal and background after imposing the improved cuts for $g'=0.01$ $\mathrm{GeV^{-1}}$ at the CEPC with $\sqrt s=240$ GeV and $\mathcal{L}=5.6$ $\mathrm{ab^{-1}}$ for the process $e^+e^-\rightarrow e^+e^-e^+e^-$.}
\label{tab5}
\begin{tabular}{c c c c c c c}\hline
  \multicolumn{7}{c}{Cross sections for signal(background) (fb)} \\ \hline
    $M_{Z_x}$ (GeV)  &~~Basic cuts ~~ & ~~Cut1~~&~~Cut2~~&~~Cut3~~&~~Cut4~~&~~SS~~\\ \hline
     10    & $0.2317$ ~&~ $0.2178$ ~&~ $0.1965$ ~&~ $0.1784$ ~&~ $0.1533$ ~~&~~10.7950~~\\
     ~     &(14.77)               &(14.18)                 &(12.30)                &(10.77)                    &(0.9745)                 & ~ \\
     20    & $0.3526$~ & ~$0.3360$~ &~ $0.3124$ ~&~ $0.2905$ ~&~ $0.2465$ ~&~10.9270~~\\
     ~     &(14.77)                     &(14.18)              &(12.30)            &(10.77)                    &(2.602)                    &~     \\
     30    & $0.4241$ ~& ~$0.4061$~ &~ $0.3765$ ~&~ $0.3495$ ~&~ $0.3027$ ~&~10.2330~~\\
     ~     &(14.77)                     &(14.18)               &(12.30)              &(10.77)                    &(4.596)                    &~ \\
     40    & $0.3876$ ~& ~$0.3698$ ~& ~$0.3323$ ~&~ $0.2984$ ~&~ $0.2674$ ~&~8.3090~~\\
     ~     &(14.77)                     &(14.11)                     &(11.27)        &(9.133)
     &(5.532)                     &~     \\
     50    & $0.3261$~ &~ $0.3108$ ~&~$0.2839$~&~ $0.2588$ ~&~ $0.2357$ ~&~6.6985~~\\
     ~     &(14.77)                     &(14.11)                     &(11.27)        &(9.133)
     &(6.696)                     &~     \\
     60    & $0.2672$~ &~ $0.2544$ ~&~ $0.2338$~ &~ $0.2148$ ~&~ $0.1984$ ~&~5.3463~~\\
     ~     &(14.77)                     &(14.11)                     &(11.27)        &(9.133)
     &(7.508)                     &~     \\
     \hline
\end{tabular}
\end{center}
\end{table}
\begin{figure}
  \centering{
  \includegraphics[width=0.6\textwidth]{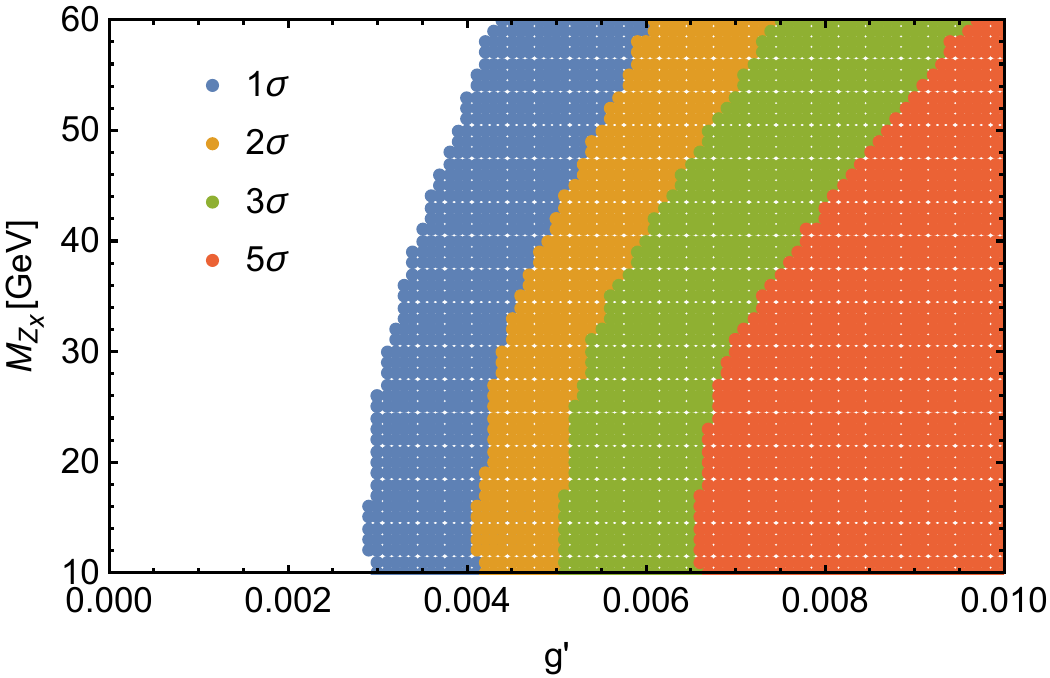}\\
  \caption{The $1\sigma$, $2\sigma$, $3\sigma$ and $5\sigma$ regions for the process $e^+e^-\rightarrow e^+e^-e^+e^-$ at the CEPC with $\sqrt s =
  240$ GeV and $\mathcal{L}=5.6$ $\mathrm{ab^{-1}}$ in the $g'$-$M_{Z_x}$ plane.}
  \label{FIG6}}
\end{figure}

\subsection{The visible decay channel $Z_x\rightarrow \nu_\ell\bar{\nu_\ell}$}

\begin{figure}
  \centering{
  \subfigure[]{\includegraphics[width=0.49\textwidth]{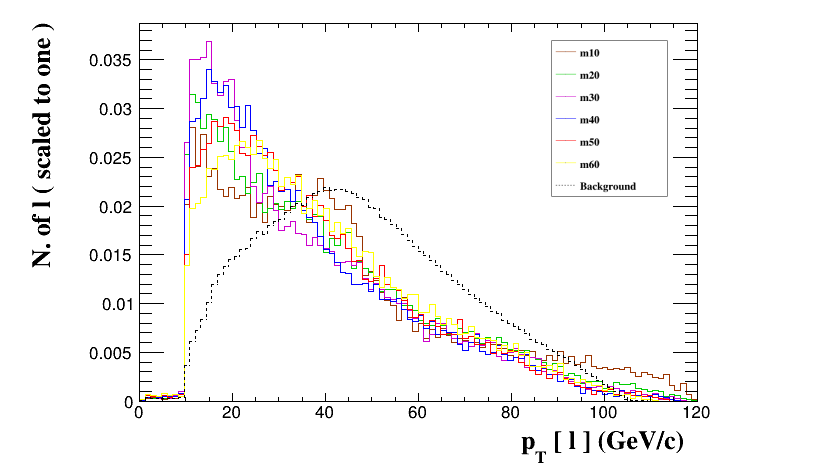}}
  \subfigure[]{\includegraphics[width=0.49\textwidth]{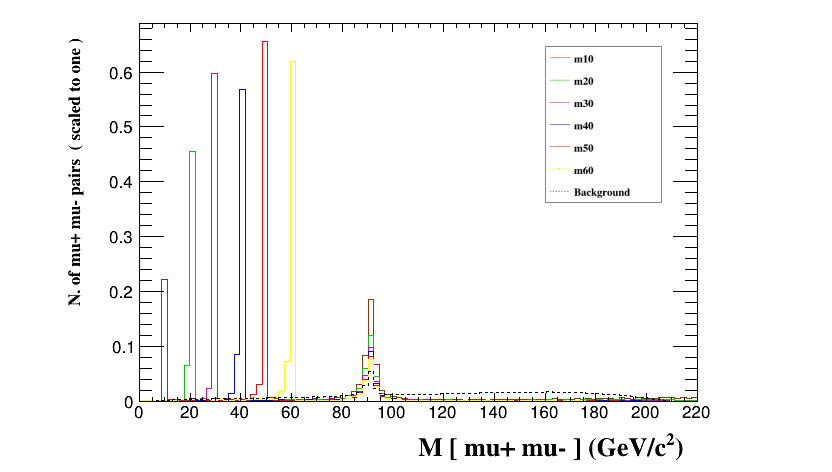}}\\
  \subfigure[]{\includegraphics[width=0.49\textwidth]{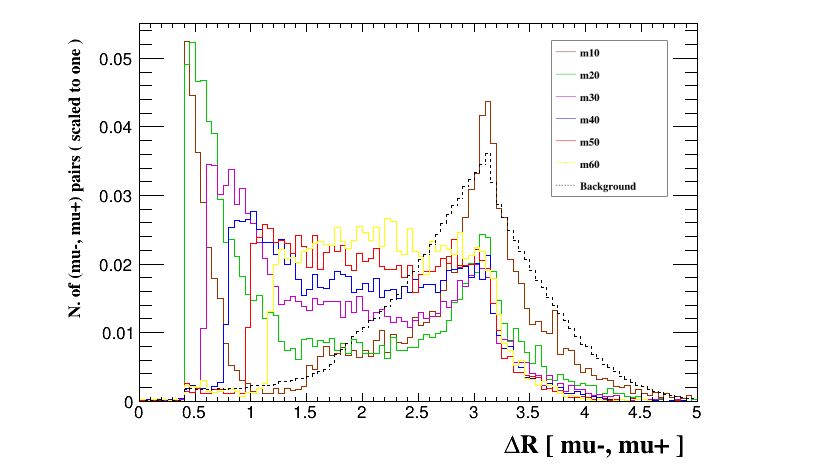}}
  \subfigure[]{\includegraphics[width=0.49\textwidth]{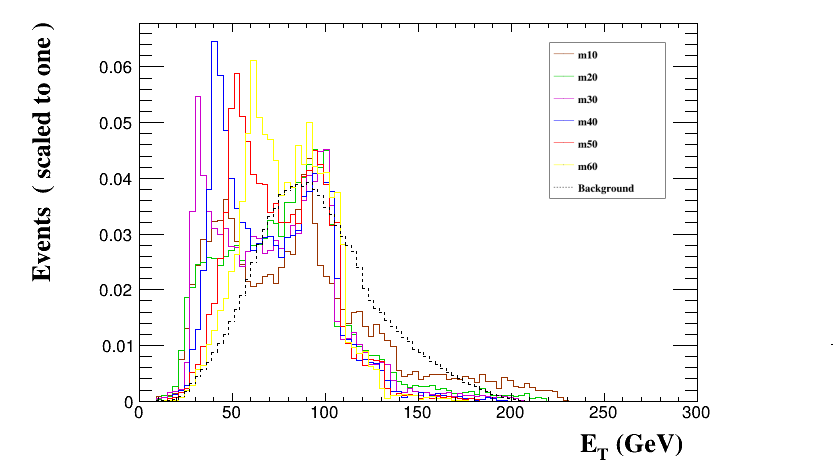}}
  \caption{Normalized distributions of $P_T(\ell)$ (a), $M(e^+, e^-)$ (b), $\triangle R$ (c) and $E_T$ (d) from the signal and background events for
  different $M_{Z_x}$ benchmark points for the process $e^+e^-\rightarrow \mu^+\mu^-\nu_\mu\bar{\nu_\mu}$ at the CEPC with $\sqrt s=240$ GeV and
  $\mathcal{L}=5.6$ $\mathrm{ab^{-1}}$.}
  \label{FIG 7}}
\end{figure}
If the $Z_x$ boson decays to a pair of neutrinos, the processes $e^+e^-\rightarrow e^+e^-\nu_e\bar{\nu_e}$ and $e^+e^-\rightarrow \mu^+\mu^-\nu_\mu\bar{\nu_\mu}$ have the larger cross sections compared to the $Z_x$ boson decays to a pair of leptons.
\begin{table}[!htb]
\begin{center}
\caption{The improved cuts for the process $e^+e^-\rightarrow \mu^+\mu^-\nu_\mu\bar{\nu_\mu}$.}
\label{tab6}
\begin{tabular}{ c c }\hline
\multirow{2}*{~~~Cut~~~}             &Mass \\ \cline{2-2}
~                          &~~~$10\mathrm{GeV}\leq M_{Z_x}\leq60\mathrm{GeV}$~~~           \\  \hline
~~~Cut1~~~             &~~~~~~$P_T(\ell)>5$~~~             \\
~~~Cut2~~~             &~~~~~~$\mid M(\mu^+, \mu^-)-M_{Z_x}\mid\leq5$~~~     \\
~~~Cut3~~~             &~~~~~~$\triangle R<4$~~~               \\
~~~Cut4~~~             &~~~~~~$E_T<150$~~~       \\ \hline
\end{tabular}
\end{center}
\end{table}
For the process $e^+e^-\rightarrow \mu^+\mu^-\nu_\mu\bar{\nu_\mu}$, according to the kinetic distributions in Figure 7,
\begin{table}[!htb]
\begin{center}
\caption{The cross sections of the signal and background after imposing the improved cuts for $g'=0.01$ $\mathrm{GeV^{-1}}$ at the CEPC with $\sqrt s=240$ GeV and $\mathcal{L}=5.6$ $\mathrm{ab^{-1}}$ for the process $e^+e^-\rightarrow \mu^+\mu^-\nu_\mu\bar{\nu_\mu}$.}
\label{tab7}
\begin{tabular}{c c c c c c c}\hline
  \multicolumn{7}{c}{Cross sections for signal(background) (fb)} \\ \hline
    $M_{Z_x}$ (GeV)  &~~Basic cuts ~~ & ~~Cut1~~&~~Cut2~~&~~Cut3~~&~~Cut4~~&~~SS~~\\ \hline
     10    & $0.1708$ ~&~ $0.1706$ ~&~ $0.03665$ ~&~ $0.03665$ ~&~ $0.03665$ ~~&~~5.4300\\
     ~     &(36.63)               &(36.59)                 &(0.2205)                &(0.2205)                    &(0.2200)                 & ~ \\
     20    & $0.2122$~ & ~$0.2215$~ &~ $0.1054$ ~&~ $0.1054$ ~&~ $0.1054$ ~&~10.1420\\
     ~     &(36.63)                     &(36.59)              &(0.4993)            &(0.4993)                    &(0.4985)                    &~
     \\
     30    & $0.2216$ ~& ~$0.2215$~ &~ $0.1317$ ~&~ $0.1317$ ~&~ $0.1317$ ~&~11.0030\\
     ~     &(36.63)                     &(36.59)               &(0.6723)              &(0.6721)                    &(0.6711)                    &~
     \\
     40    & $0.2229$ ~& ~$0.2215$ ~& ~$0.1317$ ~&~ $0.1317$ ~&~ $0.1414$ ~&~10.8980\\
     ~     &(36.63)                     &(36.59)                     &(0.8064)        &(0.8056)
     &(0.8046)                     &~     \\
     50    & $0.2205$~ &~ $0.2204$ ~&~$0.1467$~&~ $0.1467$ ~&~ $0.1467$ ~&~10.5890\\
     ~     &(36.63)                     &(36.59)                     &(0.9205)        &(0.9195)
     &(0.9195)                     &~     \\
     60    & $0.2153$~ &~ $0.2153$ ~&~ $0.1443$~ &~ $0.1443$ ~&~ $0.1443$ ~&~9.9480\\
     ~     &(36.63)                     &(36.59)                     &(1.037)        &(1.036)
     &(1.034)                     &~     \\
     \hline
\end{tabular}
\end{center}
\end{table}
\begin{figure}
  \centering
  \includegraphics[width=0.6\textwidth]{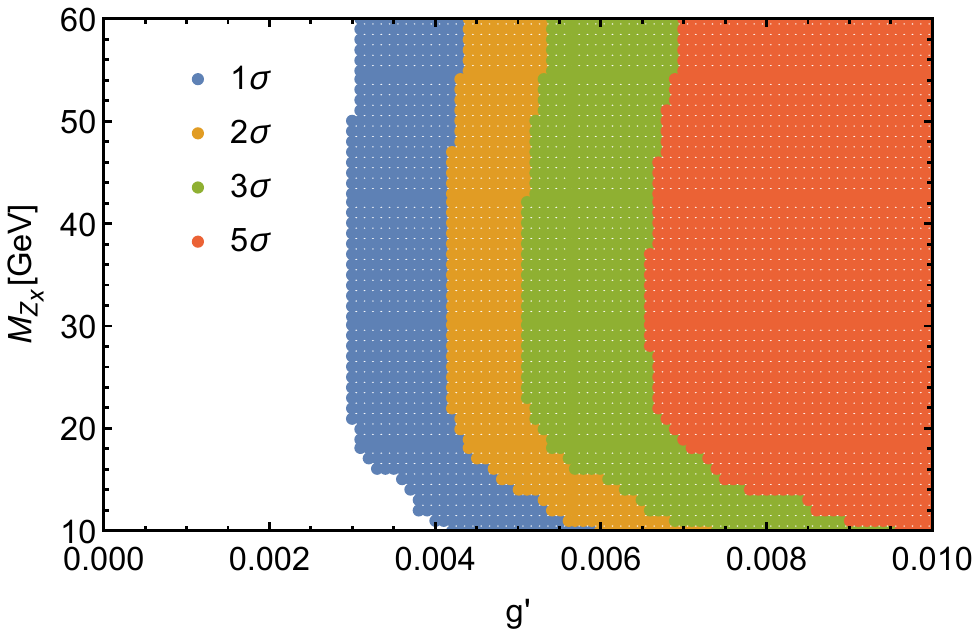}\\
  \caption{The $1\sigma$, $2\sigma$, $3\sigma$ and $5\sigma$ regions for the process $e^+e^-\rightarrow \mu^+\mu^-\nu_\mu\bar{\nu_\mu}$ at the CEPC
  with $\sqrt s = 240$ GeV and $\mathcal{L}=5.6$ $\mathrm{ab^{-1}}$ in the $g'$-$M_{Z_x}$ plane.}\label{FIG 8}
\end{figure}
the transverse momentum $P_T(\ell)$, invariant mass $M(\mu^+, \mu^-)$, angular separation $\triangle R$ between two muons and transverse energy $E_T$
are improved cuts in Table \ref{tab6} in the entire mass range $M_{Z_x}=10$ $-$ $60$ GeV. Optimized cuts might preserve as many signal events as possible, then we give the signal and background cross sections after imposing the optimized cuts for the process in Table \ref{tab7}. We can see that when the background is suppressed by two orders of magnitude, the signal is also substantially preserved. Figure 8 gives SS $=1\sigma, 2\sigma, 3\sigma, 5\sigma$ ranges in the $g'$-$M_{Z_x}$ plane, the constraints on the $Z_x$ boson are very strict with the coupling constant $g'$ reaching
$6.7\times10^{-3}$ $\mathrm{GeV^{-1}}$ at SS $= 5\sigma$.

When $Z_x$ decays to $\nu_e\bar{\nu_e}$, the peak distribution of the $e^+$ energy for the signal process $e^+e^-\rightarrow e^+e^-\nu_e\bar{\nu_e}$ is clearly demarcated from the background in Figure 9. For the low mass range $M_{Z_x}=10$ $-$ $40$ GeV, the $e^+$ energy retains more signals after applying the cuts. while on the contrary, for the large mass range $M_{Z_x}=40$ $-$ $60$ GeV, the signal events of the invariant mass $M(e^+, e^-)$ outnumber the signal events of $E(e^+)$ after improving cuts, so we add to the effective cuts at $M_{Z_x}=40$ GeV, as shown in Table \ref{tab8}. At last, Table \ref{tab9}
\begin{figure}
  \centering{
  \subfigure[]{\includegraphics[width=0.49\textwidth]{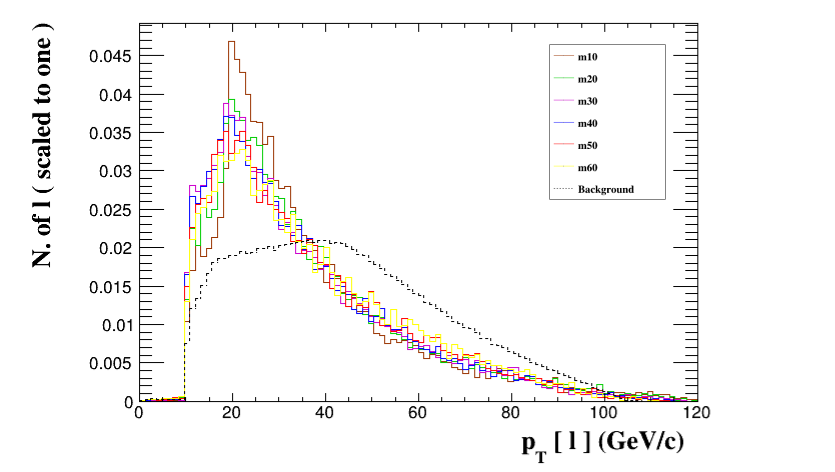}}
  \subfigure[]{\includegraphics[width=0.49\textwidth]{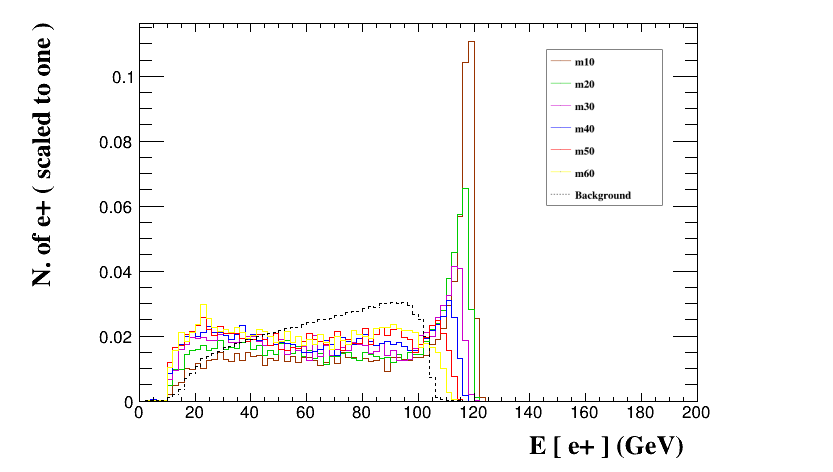}}
  \subfigure[]{\includegraphics[width=0.49\textwidth]{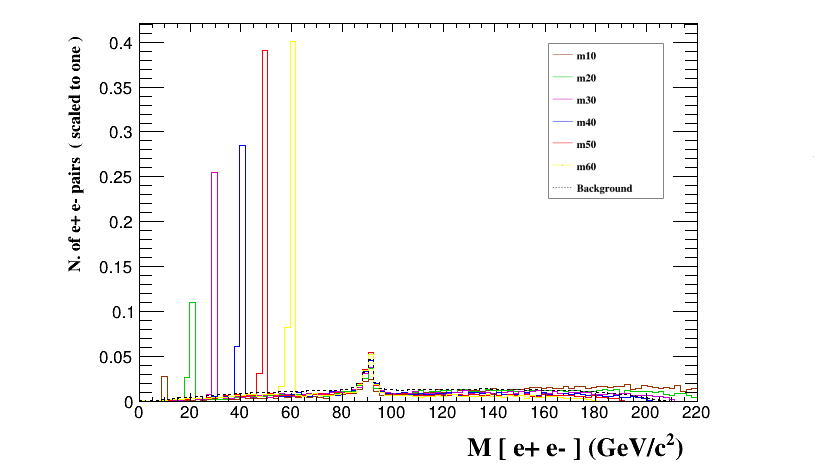}}
  \caption{Normalized distributions of $P_T(\ell)$ (a), $E(e^+)$ (b), and $M(e^+, e^-)$ (c) from the signal and background events for different
  $M_{Z_x}$ benchmark points for the process $e^+e^-\rightarrow e^+e^-\nu_e\bar{\nu_e}$ at the CEPC with $\sqrt s=240$ GeV and $\mathcal{L}=5.6$
  $\mathrm{ab^{-1}}$.}
  \label{FIG 9}}
\end{figure}
\begin{table}[!htb]
\begin{center}
\caption{The improved cuts for the process $e^+e^-\rightarrow e^+e^-\nu_e\bar{\nu_e}$.}
\label{tab8}
\vspace*{0.5cm}
\begin{tabular}{ c c c }\hline
\multirow{2}*{~~~Cut~~~}             &\multicolumn{2}{c}{Mass} \\ \cline{2-3}
~                          &~~~$10\mathrm{GeV}\leq M_{Z_x}\leq40\mathrm{GeV}$~~~    &$40\mathrm{GeV}<M_{Z_x}\leq60\mathrm{GeV}$         \\  \hline
~~~Cut1~~~             &~~~$P_T(\ell)>5$~~~       &~~~$P_T(\ell)>5$~~~          \\
~~~Cut2~~~             &~~~$E(e^+)>110$~~~       &$\mid M(e^+, e^-)-M_{Z_x}\mid\leq5$          \\ \hline
\end{tabular}
\end{center}
\end{table}
gives the cross sections of the signal and background after the improved cuts imposed on the $e^+e^-\rightarrow e^+e^-\nu_e\bar{\nu_e}$ process, and
we plot $1\sigma$, $2\sigma$, $3\sigma$ and $5\sigma$ ranges in Figure 10. The sensitivity projections of the $Z_x$ boson that we obtain are very strict for the process, especially in the region of mass $M_{Z_x}=10$ $-$ $40$ GeV, and in contrast to the three processes mentioned above, this process is more sensitive to the $Z_x$ boson.
\begin{table}[!htb]
\begin{center}
\caption{The cross sections of the signal and background after imposing the improved cuts for $g'=0.01$ $\mathrm{GeV^{-1}}$ at the CEPC with $\sqrt s=240$ GeV and $\mathcal{L}=5.6$ $\mathrm{ab^{-1}}$ for the process $e^+e^-\rightarrow e^+e^-\nu_e\bar{\nu_e}$.}
\label{tab9}
\begin{tabular}{c c c c c}\hline
  \multicolumn{5}{c}{Cross sections for signal(background) (fb)} \\ \hline
    $M_{Z_x}$ (GeV)  &~~Basic cuts ~~ & ~~Cut1~~&~~Cut2~~&~~SS~~\\ \hline
     10    & $1.6011$ ~~&~~ $1.5658$ ~~&~~ $0.49924$ ~~&~~51.6830\\
     ~     &(48.23)                     &(46.75)                     &(0.02334)                    &~     \\
     20    & $0.8491$~~ & ~~$0.8312$~~ &~~ $0.1709$ ~~&~~28.9830\\
     ~     &(48.23)                     &(46.75)                     &(0.02334)                    &~     \\
     30    & $0.5879$ ~~& ~~$0.5761$~~ &~~ $0.06803$ ~~&~~16.7940\\
     ~     &(48.23)                     &(46.75)                     &(0.02334)                    &~     \\
     40    & $0.4583$ ~~& ~~$0.4491$ ~~& ~~$0.02810$ ~~&~~9.2220\\
     ~     &(48.23)                     &(46.75)                     &(0.02334)                   &~     \\
     50    & $0.3755$~~ &~~ $0.3677$ ~~&~~ $0.1192$~~ &~~7.3630\\
     ~     &(48.233)                     &(46.75)                     &(1.522)                     &~     \\
     60    & $0.3247$~~ &~~ $0.3172$ ~~&~~ $0.1187$~~ &~~6.6180\\
     ~     &(48.23)                     &(46.75)                     &(0.1187)                     &~     \\
     \hline
\end{tabular}
\end{center}
\end{table}
\begin{figure}
  \centering
  \includegraphics[width=0.6\textwidth]{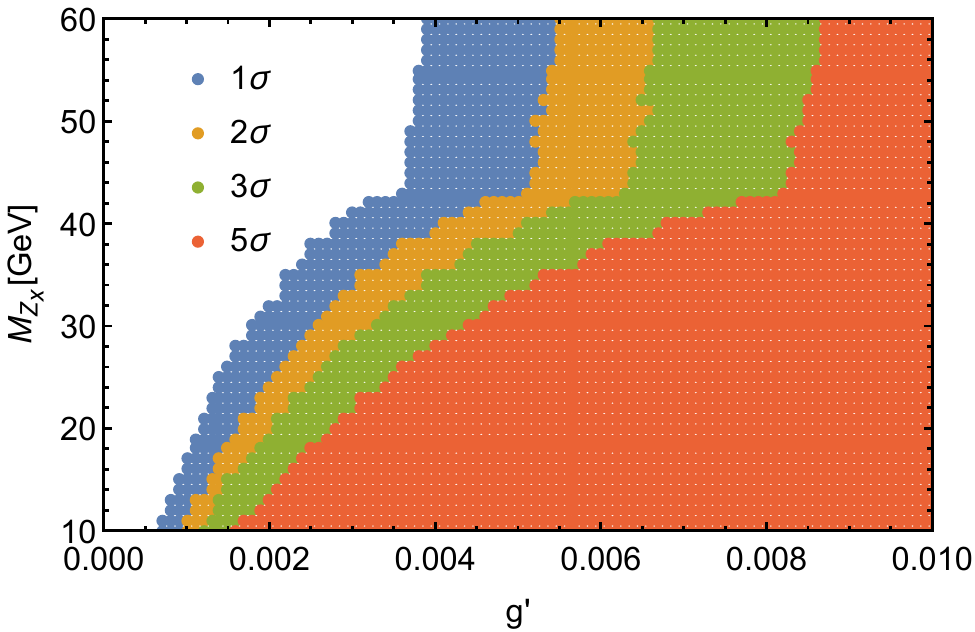}\\
  \caption{The $1\sigma$, $2\sigma$, $3\sigma$ and $5\sigma$ regions for the process $e^+e^-\rightarrow e^+e^-\nu_e\bar{\nu_e}$ at the CEPC with
  $\sqrt s = 240$ GeV and $\mathcal{L}=5.6$ $\mathrm{ab^{-1}}$ in the $g'$-$M_{Z_x}$ plane.}\label{FIG 10}
\end{figure}

\section{Conclusion and discussion}

Nowadays, there have been many works related to the leptophilic gauge boson $Z_x$, the search for the mass $10-500$ GeV $Z_x$ in the
$U(1)_{L_\mu-L_\tau}$ model is widely studied at the LHC, but the search for small mass $Z_x$ is very limited in the $U(1)_{L_e-L_\mu}$ model at the
future $e^+e^-$ colliders. It is evident that there is still a large parameter space around the electroweak scale for us to explore the $Z_x$ boson \cite{Dasgupta:2023zrh}. So we can search for the $Z_x$ predicted by the $U(1)_{L_e-L_\mu}$ model at the CEPC, as to facilitate the extension of the sensitivity of $Z_x$ or stricter couplings. In our work, we study the prospects of the CEPC to unravel NP which is associated with a new weak interaction, and the gauge boson $Z_x$ only couples to the $e$ and $\mu$ subsets in the $U(1)_{L_e-L_\mu}$ model.

We have investigated the sensitivity of the CEPC with $\sqrt s=240$ GeV and $\mathcal{L}=5.6$ $\mathrm{ab^{-1}}$ to the coupling parameter $g'$ within $M_{Z_x}=10-60$ GeV at. As can be seen from the four processes explored in the previous sections, the expected bounds of the process
$e^+e^-\rightarrow e^+e^-\nu_e\bar{\nu_e}$ on $g'$ can reach $1.0\times10^{-3}$ ($1.6\times10^{-3}$) $\mathrm{GeV^{-1}}$ for $M_{Z_x}=10-40$ GeV at $3\sigma$ ($5\sigma$), which is the most strict constraints on the $U(1)_{L_e-L_\mu}$ model. While in the $Z_x$ mass range $40-60$ GeV, the most strict constraints come from the process $e^+e^-\rightarrow \mu^+\mu^-\nu_\mu\bar{\nu_\mu}$, the expected bounds on $g'$ can reach $5.1\times10^{-3}$ ($6.7\times10^{-3}$) $\mathrm{GeV^{-1}}$ at $3\sigma$ ($5\sigma$). Compared to the other three processes, the process $e^+e^-\rightarrow \mu^+\mu^-\mu^+\mu^-$ is much looser.

In conclusion, the expected sensitivities of the four processes to parameter space of the $U(1)_{L_e-L_\mu}$ model are different. We compare our numerical results with the Figure 2 in Ref. \cite{Dasgupta:2023zrh}, they are not experimentally excluded except from the process $e^+e^-\rightarrow \mu^+\mu^-\mu^+\mu^-$. As the same time, Ref. \cite{He:2017zzr} indicates that the sensitivity to $g'$ for the process $e^+e^-\rightarrow Z_x\gamma$ can be as low as $5\times10^{-3}$ in the mass range $10-60$ GeV at $2\sigma$ level. Our results can reach $1\times10^{-3}$ within $M_{Z_x}=10$ GeV via the process $e^+e^-\rightarrow e^+e^-\nu_e\bar{\nu_e}$, and the constraints from the process $e^+e^-\rightarrow \mu^+\mu^-\nu_\mu\bar{\nu_\mu}$ can be reach $4.2\times10^{-3}$ at $2\sigma$ level, in the entire mass range $10-60$ GeV. Our numerical results are complementary to the Ref. \cite{He:2017zzr}, same conclusions are also apply to the process $e^+e^-\rightarrow e^+e^-e^+e^-$.  So, searching for the $Z_x$ boson predicted by the $U(1)_{L_e-L_\mu}$ model at the 240 GeV CEPC via processes $e^+e^-\rightarrow \ell^+\ell^-Z_x(Z_x\rightarrow \nu_\ell\bar{\nu_\ell}$ or $\ell\ell)$ can not only enhance the sensitivity projections to the parameter space, but also add explorations of future $e^+e^-$ colliders for the $U(1)_{L_e-L_\mu}$ model, which provides an another possibility to further discover the leptophilic gauge boson $Z_x$.

\section*{Acknowledgement}

This work was partially supported by the National Natural Science Foundation of China under Grants No. 11875157 and Grant No. 12147214. Yan-Yu Li would like to thank Han Wang for very useful discussions.


\bibliography{eezllll}
\end{document}